# Enhancing Critical Thinking in Generative AI Search with Metacognitive Prompts

**Anjali Singh**  The University of Texas at Austin, USA | anjali.singh@ischool.utexas.edu
**Zhitong Guan**  The University of Texas at Austin, USA | klarazt@utexas.edu
**Soo Young Rieh**  The University of Texas at Austin, USA | rieh@ischool.utexas.edu

## ABSTRACT
The growing use of Generative AI (GenAI) conversational search tools has raised concerns about their effects on people's metacognitive engagement, critical thinking, and learning. As people increasingly rely on GenAI to perform tasks such as analyzing and applying information, they may become less actively engaged in thinking and learning. This study examines whether metacognitive prompts—designed to encourage people to pause, reflect, assess their understanding, and consider multiple perspectives—can support critical thinking during GenAI-based search. We conducted a user study (N=40) with university students to investigate the impact of metacognitive prompts on their thought processes and search behaviors while searching with a GenAI tool. We found that these prompts led to more active engagement, leading students to explore a broader range of topics and engage in deeper inquiry through follow-up queries. Students reported that the prompts were especially helpful for considering overlooked perspectives, promoting evaluation of AI responses, and identifying key takeaways. Additionally, the effectiveness of these prompts was influenced by students' metacognitive flexibility. Our findings highlight the potential of metacognitive prompts to foster critical thinking and provide insights for designing and implementing metacognitive support in human-AI interactions.

## INTRODUCTION
Around half of the individuals in the U.S. aged 14–22 now use Generative AI (GenAI) to seek information, as found by a recent study (Teen and Young Adult Perspectives on Generative AI, 2024). The ability of GenAI tools to synthesize vast amounts of online information into well-structured and fluent responses is helpful for streamlining the search process, and reduces the effort required compared to using traditional search engines. Further, tools such as Perplexity.ai and Google AI Overview provide access to the sources from which information is retrieved, enabling users to verify responses and access additional information. However, the increased convenience offered by GenAI tools may come at a cost. Traditional search engines necessitate active seeking and verification of sources (Shah & Bender, 2024). Research on search as learning (SAL) has emphasized the importance of learning occurring during the search process, as users engage in a range of cognitive activities such as understanding, applying, analyzing, and evaluating information (Jansen et al., 2009; Rieh et al., 2016). In contrast, rapid access to synthesized responses from GenAI could potentially bypass these cognitive processes, leading to passive consumption of information (Venkit et al., 2024) and less exposure to diverse perspectives (Sharma et al., 2024).

Beyond these concerns, GenAI tools impose metacognitive demands on users (Tankelevitch et al., 2024). Metacognition—the *awareness* and *regulation* of one's thinking (Winne, 2017)—is essential for interacting with GenAI tools. These tools require verbalized prompting, which requires self-awareness of task goals and task planning. Moreover, while GenAI responses require validation due to potential inaccuracies, they often appear confident, lacking explicit representations of uncertainty or the ability to convey their absence (Kidd & Birhane, 2023). This potentially increases their persuasiveness, which can hinder their careful evaluation. As Tankelevitch et al. argue, critical evaluation of GenAI responses requires well-calibrated confidence in one's knowledge of the search topic, along with metacognitive flexibility to refine prompting strategies as needed. This can be particularly challenging for users who either lack sufficient knowledge of a subject or exhibit misplaced confidence in their level of expertise on a given topic (Singh et al., 2025).

Prior research on metacognition has explored the use of *metacognitive prompts* to support learning in educational settings (Bannert & Mengelkamp, 2013; Lin & Lehman, 1999) as well as to facilitate searching with traditional search engines (Hwang & Kuo, 2011; Stadtler & Bromme, 2008). Metacognitive prompts focus people's attention on their own thoughts and on understanding the activities in which they are engaged during learning (Lin, 2001). Their use has been shown to significantly enhance the effectiveness of information seeking in traditional search processes (Zhou & Lam, 2019). However, their effectiveness in influencing search behavior and promoting critical thinking during interactions with GenAI tools remains unexplored.

This study investigates the effects of metacognitive prompts on search behaviors and critical thinking during searching with the GenAI tool Perplexity (https://www.Perplexity.ai/). We conducted a between-subjects user study with 40 students from the University of Texas at Austin. Participants were randomly assigned to one of two conditions: the *Cues* condition, in which they received metacognitive prompts during searching, or the *Baseline* condition, in which no prompts were provided. Within the context of using GenAI tools to search for information, this study examines the following research questions:

**RQ1.** What are the differences in students' search behaviors, demonstrated critical thinking, and perceived impact on thinking and searching, when searching with versus without metacognitive prompts?
**RQ2.** What are the perceived effects of each metacognitive prompt on students' search behavior and thinking?
**RQ3.** To what extent do students perceive each metacognitive prompt as helpful following the search session?

## RELATED WORK

### Search as Learning

The Search as Learning (SAL) framework connects search behavior, learning processes, and learning outcomes (Rieh et al, 2014; Eickhoff et al., 2017). SAL researchers have reconceptualized search systems as rich, interactive learning environments where learning occurs over the course of search sessions (Rieh et al., 2016). SAL studies aim to better support learning during search, also emphasizing learning as a primary outcome of the search process (Hansen & Rieh, 2016). According to Vakkari (2016), meaningful learning requires more than high-quality search results: it requires presenting information in ways that help users integrate it into their existing knowledge structures. Rieh et al. (2016) identified various search activities associated with critical and creative learning modes, including comparing, differentiating, sense-making, assessing credibility, and evaluating usefulness. More recently, Urgo and Arguello (2025) argued that facilitating self-regulated learning, in which learners understand and control their learning process (Zimmerman, 2002) during search, is crucial for future research. The Search as Learning (SAL) framework offers a valuable foundation for investigating metacognition in the context of generative AI search, as it emphasizes the multiple phases of the search process such as task definition, goal setting and planning, strategy and tactic selection, and adaptation.

### Searching with GenAI Tools

GenAI search tools, which aggregate and synthesize information from multiple sources, challenge traditional search engines by offering greater convenience and speed (Zhou & Li, 2024). While these tools enhance user engagement and experience through interactive dialogue, ability to handle complex queries and generating personalized responses, they also risk presenting inaccurate or misleading information with undue confidence (Amer & Elboghdadly, 2024). This shift in information search and response presentation by GenAI tools influences how users interact with and process information for learning. In particular, the widespread adoption of GenAI tools has sparked concerns about "cognitive offloading" (Risko & Gilbert, 2016), i.e., when users delegate cognitive tasks to AI, reducing their own cognitive engagement and impacting their ability to critically engage with information (Fan et al., 2024). This issue is particularly pronounced among younger users of GenAI tools, such as college students (Gerlich, 2025). Interactions with GenAI tools can also lead to selective attention and retention. A recent study found that users favored consonant over dissonant opinions while interacting with GenAI search tools (Sharma et al., 2024). Additionally, the populist nature of GenAI search, which prioritizes mainstream perspectives while sidelining alternative viewpoints, can contribute to informational homogenization (Amer & Elboghdadly, 2024; Solaiman et al., 2023). To address these issues, this work explores whether metacognitive prompts that support searchers' self-regulated learning can help them to engage with GenAI search tools more actively and critically.

### Metacognition and Critical Thinking

Metacognition refers to the awareness and understanding of one's own cognitive processes. It involves the ability to monitor, evaluate, and regulate one's thinking and learning strategies (Flavell, 1979). Metacognitive abilities consist of: (i) metacognitive *monitoring*: the ability to assess one's understanding and confidence, as well as the capacity to adjust these judgements as needed, and (ii) metacognitive *control:* the regulation and adjustment of cognitive processes to improve learning, such as changing strategies when encountering difficulties (Efklides, 2008; Nelson & Narens, 1990). Metacognition is essential for critical thinking, which, in turn, is essential for deep and effective learning (Magno, 2010). While definitions of critical thinking vary across disciplines (Beyer, 1984; Mayer & Goodchild, 1990; Halpern, 2013), we adopt Dewey's conceptualization of it as reflective thinking (Dewey, 1910), due to its structured and process-oriented nature. Dewey defined it as conscious evaluation of evidence, occurring when the basis of a belief is carefully examined. He elaborated that it involves active, persistent consideration of ideas, openness to and tolerance for uncertainty during inquiry, and the ability to connect and evaluate related ideas. Metacognition improves critical thinking by enabling reflection, assessment, and adjustment of thought processes to improve learning (Ku & Ho, 2010).

Yet many people struggle to engage in metacognitive activities spontaneously (Bannert & Mengelkamp, 2013), therefore providing effective metacognitive support could help them achieve their learning goals. A promising approach is the use of *metacognitive prompts*, which direct an individual's attention to their own thought processes and the learning activities in which they are engaged (Lin, 2001; Lin & Lehman, 1999). These prompts, typically delivered in the form of guiding questions, offer explanations, reminders, or hints about what learners are doing and how and why they are doing it (Berardi-Coletta et al., 1995). They are particularly useful for those who possess metacognitive skills but may not be able to apply them spontaneously, such as college students (Bannert & Mengelkamp, 2013).

Prior research has shown that metacognitive strategies, rather than cognitive skills alone, enhance search efficiency (Hwang & Kuo, 2011; Stadtler & Bromme, 2008). Several studies have explored the use of metacognitive prompts to activate search strategies, such as to assist searchers in finding information systematically (Stadtler & Bromme, 2007) and to direct people to evaluate the quality of potential sources (Leeder & Shah, 2016). The use of metacognitive prompts has been significantly associated with improvements in online search processes (Zhou & Lam, 2019). This highlights their potential to enhance search behavior among users engaging with GenAI tools, particularly in fostering critical thinking about the search topic and the AI-generated responses.

**RESEARCH METHODS**

This research adopts a between-subjects design using surveys and retrospective think-aloud interviews to examine the impact of metacognitive prompts (hereafter referred to as "cues") on students' critical thinking and search behaviors while searching with the GenAI tool, Perplexity. We selected Perplexity due to its widespread use as a conversational AI search tool that provides synthesized responses with consistent source citations. Unlike other GenAI tools, such as ChatGPT, it allows users to access and explore sources underlying the generated information. Further, its interface facilitates step-by-step information exploration (Zamani et al., 2023).

To simulate a system capable of delivering cues to support GenAI based search, we employed the Wizard-of-Oz method (Maulsby et al., 1993), in which a researcher mimicked system-generated cues by sending them at strategically timed moments within an authentic search context. Given the absence of an existing system with such capabilities, this approach was selected to inform future system development. The between-subjects design enabled us to experimentally compare two conditions (with versus without cues) to assess the cues' effects on students' search behaviors and critical thinking (RQ1). A mixed methods approach was chosen to capture both the measurable outcomes and the underlying reasoning behind participants' actions. Quantitative data from search interactions and post-task questionnaires provided insights regarding participant behavior and thinking (RQ1) and perceptions, while qualitative data from surveys and think-aloud interviews revealed how cues influenced students' thinking and search behavior (RQ2) and their perceptions of the cues' helpfulness (RQ3).

**Study Design**

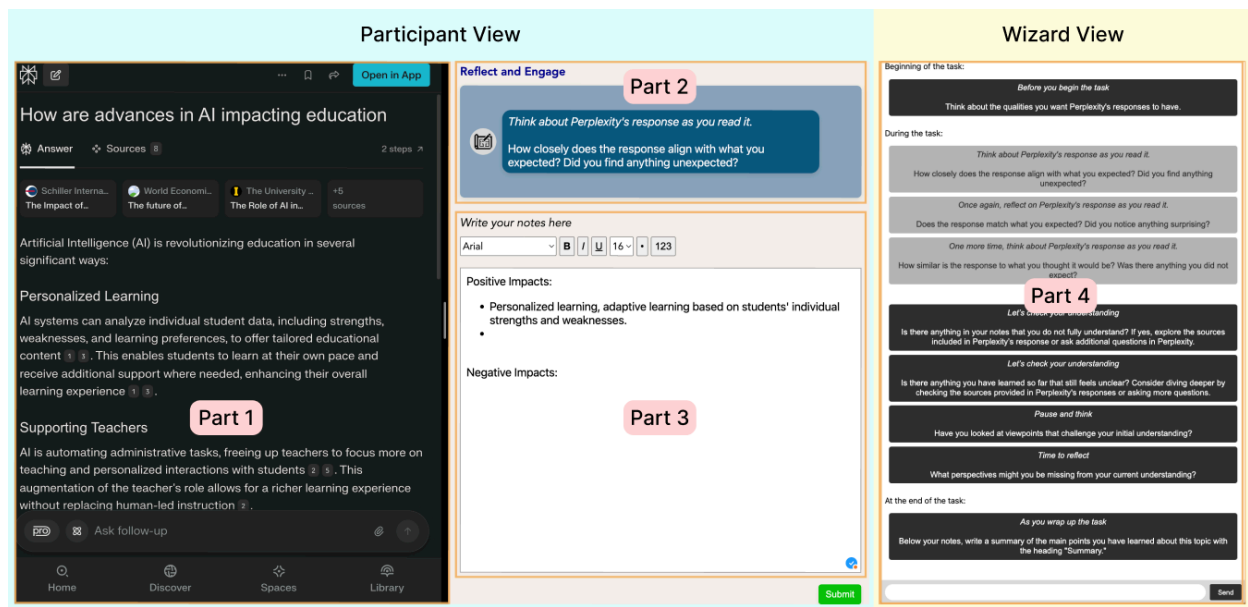

Figure 1. Interfaces used in the study: Participants were instructed to arrange Perplexity (Part 1) and the note-taking interface (consisting of only Part 3 for the *Baseline* condition and Parts 2 and 3 for the *Cues* condition) in a split-screen view. Part 4 shows the interface used by the researcher to deliver the cues.

The study design was informed by results from four pilot tests, which led to refinement of the wording and delivery heuristic of the cues and confirmed the suitability of a between-subjects experiment. Accordingly, a between-subjects (*n*=40) study was conducted to examine how the cues influence students' search behaviors and critical thinking during GenAI-supported search tasks. Participants were randomly assigned to one of two conditions: *Baseline* vs. *Cues*. For each condition, we developed a browser-based note-taking interface (see Figure 1) that participants were instructed to use during their search session. To simulate authentic multitasking behaviors, participants were instructed to split their screens horizontally, positioning Perplexity on the left for information search (Part 1) and their designated note-taking tool on the right. The *Baseline* interface provided a standard notepad (Part 3), while the *Cues* interface additionally integrated the cues (Part 2). In the *Cues* condition, researchers monitored participants' real-time behaviors via Zoom and sent cues from the wizard's view interface (Part 4). The *Baseline* condition offers a strong point of comparison, as note-taking inherently supports cognitive and metacognitive engagement (Makany et al., 2009) and supports sense-making in SAL contexts (Roy et al., 2021).

*Study Procedure*
The study procedure, illustrated in Figure 2, was approved by the IRB of the authors' institution. Each participant took part in an individual remote session conducted via Zoom on their laptop, lasting between 50 and 75 minutes. The session began with an overview of the study, followed by participants reviewing and signing an informed consent form. They then completed a pre-task questionnaire. Next, participants watched a tutorial video that introduced their assigned tool and explained how to split the screen to view the tool and Perplexity side by side. Following this, they received detailed task instructions, including the assigned search topic. Next, they rated their familiarity with and interest in the assigned topic. Then, they proceeded to the search task, during which they took notes in their assigned tool. The entire 25-minute session was screen-recorded for analysis.

Throughout the session, the researcher took observational notes on participants' behaviors and interactions to inform the retrospective think-aloud interviews. In the *Cues* condition, the researcher also sent cues using the Wizard view interface. The session concluded with a retrospective think-aloud interview lasting between 10 and 20 minutes, in which participants reflected on their search and note-taking process. This method was chosen to reduce the additional cognitive load associated with concurrent think-aloud protocols (Elling et al., 2011). Participants completed tasks silently to maintain natural interaction. Researchers then replayed the recorded search session via Zoom, pausing at key interactions, including submitting queries, clicking on sources, and responding to the cues—to elicit reflections through a semi-structured interview.

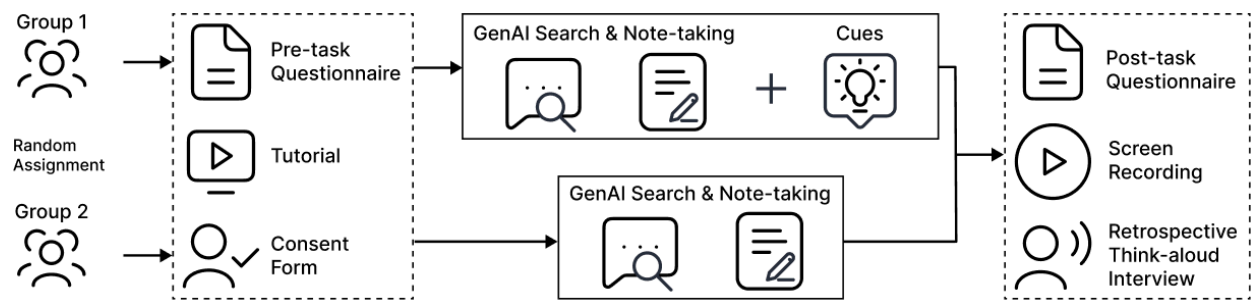

**Figure 2. Study procedure, including: (i) random assignment, (ii) consent, pre-task questionnaire, and tutorial of assigned tool, (iii) search task with the assigned note-taking tool, and (iv) post-task questionnaire followed by retrospective think-aloud interview using screen recording of search session.**

*Search as Learning Task*
The search topic was developed based on three primary criteria. It needed to be: (i) open-ended to encourage exploration of diverse perspectives, (ii) easily understandable to ensure participant engagement, and (iii) within the research team's domain of expertise for effective analysis of participant behaviors. Accordingly, we selected the topic: "How are advances in AI impacting education." Participants were given the following instructions: "*Imagine you are preparing to write an essay on this topic. Your task is to conduct research on this topic using Perplexity and take notes using <assigned tool>. As mentioned in the tutorial, consider you are taking notes for your own reference later on to help you write your essay. As you work on this task, focus on: (1) gaining a complete understanding of the topic, and (2) gathering evidence from reliable sources. Please do not use any other websites, such as Google, to search for information.*" Additionally, participants in the *Cues* condition were informed: "*To support you with this task, you will receive system-generated 'prompts' to help with your thinking process.*"

**Design of Cues**
The cues were designed to enhance students' engagement and critical thinking during GenAI-supported search and learning activities. Drawing on prior research on the metacognitive demands and affordances of GenAI tools (Tankelevitch et al., 2024) and GenAI-assisted search (Sharma et al., 2024; Venkit et al., 2024), each cue type was designed to target specific phases of the SAL process (Rieh et al., 2016). We developed five distinct categories of cues: Orienting, Monitoring, Comprehension, Broadening Perspectives, and Consolidation. Table 1 summarizes these cue types, their intended functions, and provides representative examples. To maintain participant engagement and avoid redundancy, we created two to three variations for each cue type, except for Orienting and Consolidation cues, which had only one version each.

| Type of Cue | Intended Functions | Cue Example |
|---|---|---|
| Orienting | Set evaluative criteria for GenAI outputs | *"Think about the qualities you want Perplexity's responses to have."* |
| Monitoring | Encourage comparison of GenAI outputs with prior knowledge and expectations; reduce passive acceptance | *"How closely does the response align with what you expected? Did you find anything unexpected?"* |
| Comprehension | Identify gaps in comprehension; encourage active engagement with sources and persistent questioning | *"Is there anything in [your notes/the perplexity response] that you do not fully understand? If yes, explore the sources included in Perplexity's response or ask follow-up questions."* |
| Broadening Perspectives | Encourage consideration of alternative or overlooked perspectives | *"What perspectives might you be missing from your current understanding?"* |
| Consolidation | Synthesize knowledge and reflect on what has been learned | *"Write a summary of the main points you have learned about this topic."* |

**Table 1. Cue Types, their Intended Functions, and Representative Examples**

*Cue Delivery*

Cues were delivered to the participants using the Wizard-of-Oz technique to simulate an adaptive system. To maintain consistency and minimize potential bias, cues were administered according to predefined criteria based on observable participant behaviors. An Orienting cue was sent at the beginning of the search session. A Monitoring cue was sent up to three times immediately following the generation of a GenAI response. Comprehension cues were sent up to twice per session when participants paused during reading or note-taking (indicated by inactivity in scrolling or typing), typically after the first five minutes of searching. If no pauses occurred within the first 10 minutes, these cues were sent when participants refrained from clicking on sources or asking follow-up questions. Broadening Perspectives cues were sent up to twice when participants showed excessive focus on a single viewpoint, as indicated by their queries and notes, and were typically triggered when they moved the cursor to the text box to enter a new query. Consolidation cues were delivered when participants indicated that they had completed the task or when four minutes remained in the 25-minute search session, whichever occurred first. To prevent cognitive overload, a minimum interval of 2–3 minutes was maintained between any two consecutive cues.

**Pre- and Post-Task Questionnaires**

The pre-task questionnaire assessed participants' frequency of using GenAI tools (such as ChatGPT and Gemini) and Perplexity on a 5-point Likert scale. Further, it measured their metacognitive awareness using the Metacognitive Awareness Inventory (Schraw & Dennison, 1994). After getting introduced to the search topic, participants rated their interest in and familiarity with the topic on a 5-point Likert scale. Post-task, participants rated on a 5-point Likert scale whether the assigned note-taking tool: (i) helped identify misconceptions, (ii) challenged strong beliefs about the search topic, and (iii) influenced future use of GenAI search tools. Those in the Cues condition also rated the helpfulness of each type of cue on a 5-point Likert scale.

**Data Analysis**

*Quantitative Analysis*

To systematically analyze participants' search behavior and critical thinking, the two authors who led the study sessions discussed their observations. Drawing on these insights and prior literature on SAL (Rieh et al., 2016) and critical thinking (Dewey, 1910), the first author developed an initial codebook with quantitative measures. Both authors independently coded 25% of the data and refined the codebook through discussion of disagreements. A second round of independent coding on another 25% of the data yielded an intraclass correlation coefficient (Weir, 2005) above 0.9 for each measure, indicating strong inter-rater reliability. The finalized codebook was then applied by the first author to code the entire dataset.

The final set of measures capturing search behavior were as follows (see Appendix for more details):

- *Queries*: Number of search queries used to prompt Perplexity.
- *# Sources*: Number of sources in AI-generated responses that participants clicked on.

- *# Topics*: Number of unique and relevant topics, i.e., "information focus" of one or more queries (Jansen et al., 2007) addressed through the search queries, excluding the provided search topic. A single query could consist of multiple topics, with distinct perspectives (e.g., pros and cons) counted separately.
- *# Receptive queries*: Number of unique and relevant queries focusing on understanding or comprehension, e.g., "What are some ethical considerations when integrating AI into education?"
- *# Critical queries*: Number of unique and relevant queries aimed at justifying or challenging opinions, arguments, gathering evidence, predicting, or digging deeper into a topic, e.g., "Should students be allowed to use AI for assignments?" and "Are there challenges with AI replacing human qualities in education?"

The measures capturing critical thinking were as follows:

- *Persistent Inquiry*: Whether a participant pursued an idea through one or more relevant follow-up queries.
- *Source Engagement*: Whether a participant actively engaged with sources to get more details to support the AI response, test hypotheses, or validate the accuracy of the AI response.
- *Independent Thinking*: Whether a participant added new inputs in notes in addition to information provided by the AI responses or the sources cited in them.

Given the small sample size and non-parametric distribution of participant background and search behavior measures, the Mann-Whitney U test was used for between-condition comparisons of these measures. For measures of critical thinking, which were categorical variables indicating the presence or absence of evidence, the Chi-square test of independence was used given its suitability for categorical data analysis. Following convention in psychology studies (Pritschet et al., 2016), we consider $p<0.05$ to be statistically significant, and a *p*-value between 0.05 and 0.1 to be of marginal significance, indicating a trend that is close to statistical significance.

*Thematic Analysis*
Thematic analysis was chosen to analyze the data from the think-aloud interviews and post-task reflections in order to identify common patterns in participants' thoughts and behaviors. Based on the interview transcripts, two authors employed thematic analysis to develop a codebook collaboratively, with iterative reviews and discussions among researchers to identify and define the codes, incorporating representative quotes. The same coding scheme was applied to data from both the *Cues* and *Baseline* conditions, except for some codes that were developed specifically for the *Cues* condition to capture participants' reactions to the cues. Accordingly, the following themes were applied to analyze the transcripts: (i) rationale for participants' search queries, (ii) reasons for engaging or not engaging with sources in AI responses, (iii) comprehension of AI-generated responses, (iv) note-taking approaches, and (v) for the *Cues* condition only, the influence of each cue on participants' thinking and search behavior.

**RESULTS**

**Background of Participants**
Forty students, recruited via institutional mailing lists and flyers from the University of Texas at Austin, participated in the study and were compensated $20. Recruitment continued until code saturation was achieved in the qualitative data. The final sample included 33 undergraduate and 7 master's students across 17 majors, including Business, Mathematics, Psychology, Engineering, Biology, and Computer and Information Sciences. Participants self-reported their gender as male ($n=16$), female ($n=22$), and non-binary ($n=1$), with one participant preferring not to disclose. Their ages ranged from 18 to 29 years: 18–20 ($n=21$), 21–23 ($n=16$), and 24–29 ($n=3$). Mann-Whitney U tests revealed no significant differences between the two conditions in GenAI tool use frequency (U=207.0, $p=0.850$), Perplexity use frequency (U=223.5, $p=0.514$), familiarity with the search topic (U=209.0, $p=0.789$), interest in the search topic (U=197.00, $p=0.943$), and metacognitive awareness (U=219.50, $p=0.605$). In each condition, 10% of participants reported high familiarity with the topic, 60–65% moderate familiarity, and 25–30% slight familiarity.

**Between-group Differences in Search Behavior and Critical Thinking (RQ1)**

|  | **Search Duration (minutes)** | **# Queries** | **# Sources** | **# Topics** | **# Receptive Queries** | **# Critical Queries** |
|---|---|---|---|---|---|---|
| ***Cues** (M, SD)* | 18.23, 4.59 | 6.20, 2.67 | 3.75, 2.77 | 4.90, 2.34 | 3.00, 1.49 | 1.70, 1.42 |
| ***Baseline** (M, SD)* | 16.34, 6.88 | 4.25, 3.31 | 2.75, 3.19 | 2.60, 2.28 | 2.05, 2.14 | 1.00, 1.38 |
| **Mann-Whitney U** | 218.00 | 271.50 | 254.00 | 301.00 | 267.00 | 258.00 |
| ***p*-value** | .636 | .053* | .144 | .006** | .068* | .100* |

Table 2. Mann-Whitney U Test Results for Search Behavior Measures (** indicates significance; * indicates marginal significance)

Table 2 presents the mean and standard deviation values for measures capturing search behavior. We found that participants in the Cues condition spent more time searching, created more queries, clicked on more sources, covered more topics, and asked more critical questions. Of these, the difference in the number of topics was significant (U=301.00, p=0.006). The differences in the number of queries (U=271.50, p=0.053), receptive queries (U=267.00, p=0.068), and critical queries (U=258.00, p=0.100) were marginally significant.

|  | **Persistent Inquiry** | **Source Engagement** | **Independent Thinking** |
|---|---|---|---|
| ***Cues*** *(% of participants)* | 80% | 75% | 55% |
| ***Baseline*** *(% of participants)* | 50% | 60% | 35% |
| *$X^2$ (df, N)* | 3.96 (1, 40) | 1.90 (1,4 0) | 2.51 (1, 40) |
| **Effect Size (Cramér's V)** | .31 | .22 | .25 |
| *p-value* | .047** | .168 | .113 |

**Table 3. Chi-square Test Results for each Critical Thinking Measure (** indicates significance).**

Table 3 reports: (i) percentage of participants in each condition who demonstrated critical thinking based on persistent inquiry, source engagement, and independent thinking, and (ii) results of Chi-square tests examining the association between condition and critical thinking measures. We found that more participants in the Cues condition demonstrated persistent inquiry, source engagement, and independent thinking. Of these, the association between condition and persistent inquiry was statistically significant ($\chi^2$(1,N=40) = 3.96, p=0.047, Cramér's V=0.31).

Post-task questionnaire responses assessing the perceived impact of the assigned note-taking tool on search behavior and critical thinking revealed the following mean (SD) values for each condition, with higher scores indicating greater agreement: (i) helped identify misconceptions – *Baseline*: 2.60 (1.35), *Cues*: 2.75 (1.55); (ii) challenged strong beliefs – *Baseline*: 2.45 (1.19), *Cues*: 2.95 (1.23); (iii) influenced future use of GenAI tools – *Baseline*: 2.65 (1.34), *Cues*: 2.95 (1.43). None of these differences were statistically significant.

We now turn to the retrospective think-aloud data to better understand differences in participants' thought processes and search behaviors. Although the difference in search duration across conditions was not significant (see Table 2), more participants in the *Baseline* condition ended their search early. Specifically, 6 out of 20 *Baseline* participants concluded their search within 10 minutes, compared to only 1 in the *Cues* condition. Four of these *Baseline* participants submitted only one query, which was the given topic. When asked why they stopped searching early, these participants overestimated the sufficiency of the initial AI-generated response to arrive at a comprehensive understanding of the given topic. This perception was influenced by their view that the AI-generated responses tended to be generic. One *Baseline* participant noted: "I completed my search, because whatever basic info I was looking for, I got it... I felt like, I got certain points that are very basic and very broad. Everybody can relate to it, of course" (P4). For this reason, even a participant in the *Cues* condition who ended their search early acknowledged limited engagement with sources, stating: "Even though it was a new perspective, I didn't feel the need to click on [a source] because [the response] was very general, so it looked more truthful than misinformation" (P11).

When AI responses aligned with participants' expectations, they tended to accept the information as valid, demonstrating confirmation bias. This pattern was more pronounced among participants in the *Baseline* condition. As one of them noted: "I think I came in knowing a little bit more about the topic. So, because I had assumptions about it and the responses to the questions were in line with what I already knew or predicted, I didn't think, oh, maybe this might be wrong, because it lined up with some of the bias I already had at the beginning" (P14). This may explain why participants in the *Baseline* condition posed fewer queries than those in the *Cues* condition (see Table 2). As discussed in the next section, the Monitoring, Comprehension, and Perspective Broadening cues encouraged participants in the *Cues* condition to reflect on their understanding and reset their expectations.

Across both conditions, participants' confidence in their prior knowledge or their preconceptions of Perplexity's abilities influenced both their search strategies and search duration. For example, one *Baseline* participant, who prompted Perplexity with only one query, said: "I feel that because I have been using AI tools quite a bit for my own benefit, I have a good understanding of the topic at hand… The first explanation by Perplexity itself was able to give me a few points... And I felt that was enough information for me to understand how advances in AI are impacting education since I am probably one of the persons getting impacted" (P2). For this reason, a few participants in the *Cues* condition also ended their search early, despite receiving cues that encouraged deeper engagement. One such participant explained: "I thought that [Perplexity] was kind of giving me repetitive answers… I guess it doesn't really remember what I've asked, so it's hard for me to really dive deep into detail to ask something" (P1).

**Perceived Effects of Cues on Thought Process and Search Behavior (RQ2)**
We now present findings from retrospective think-aloud interviews with the 20 *Cues* condition participants on the effects of metacognitive prompts on their thought processes and search behavior while using a GenAI tool (RQ2).

The **Orienting cue** served its intended purpose by helping most participants set their expectations for the AI responses. As a result, many of them included specific criteria into their queries, such as the desired level of detail, preferred sources for generating responses, and the inclusion of factual information, including statistical evidence: "I think when I search for something, I'm pretty blank, but having a prompt like this would encourage me to think more about what I'm looking for in a response" (P27), "[Because of this cue, I included in my query] give sources that should be independent from political influences, causes, and general blogs. Focus more on educational papers" (P3). However, some participants expressed confusion about the cue's intent or were unsure how to set expectations. One participant did not find it useful as they were already aware of the importance of query construction.

The **Monitoring cue** prompted several participants to focus on novel or unexpected information in the AI-generated response, as intended: "It was helpful because while I was reading [an AI response], I didn't necessarily think about teachers' perspectives at all. I was solely thinking of my perspective as a student, and so that kind of helped me to ask more questions..." (P19). It also encouraged some to reconsider their understanding of the given topic. One participant reflected on the structure of the AI response: "...where I had mine in negative versus positive effects [of AI in education], they didn't really do it like that. They just had different effects and kind of let the users interpret [them as] positive or negative" (P9). Others reported that the cue nudged them to verify the AI response and seek additional evidence or specific details: "[Because of this cue,] I was hoping [the AI response] would have a little more, I guess, concrete numbers" (P15), "I wished that [Perplexity] focused on negative impacts too, because I think in the initial search they mainly just said positive traits" (P37). For a few participants, the cue did not serve its intended purpose as they did not find the AI response surprising, possibly because they found it relatable: "I was trying to find something unexpected, but after reading what Perplexity told me, there wasn't really anything that was super surprising. I guess everything that I found out was something that could be reasonably assumed" (P17).

For the **Comprehension cue**, several participants reported that it encouraged them to dig deeper and seek details to support the AI responses, by asking follow-up questions and engaging with sources: "I read it and I was like, oh, I might need more information to go into more detail about whatever I was writing." (P21). Further, it encouraged them to verify the AI responses: "Perplexity was telling me the answers, but [the cue] had me thinking of how did those answers pop up. So, I think at this point it reminded me to check the sources" (P17). One participant reported actively revising their notes in response to this cue, adding clarifications and supplementary details to ensure their future self could understand the intended meaning. However, a few participants found this cue less helpful, citing reasons similar to those expressed regarding the monitoring cue—the AI-generated responses were generic or repetitive, and therefore easy to comprehend: "There wasn't anything so unclear because I think repetitively the AI was giving me similar answers" (P1).

The **Broadening perspective** cue received the most positive responses from participants. It prompted several participants to pause and reflect on overlooked perspectives, often by reviewing their notes: "At this point, I was starting to think I've already gotten the advantages. I haven't really explored the disadvantages, so I thought I'd take this time to look at challenges and any disadvantages so that I'm not looking from just a purely positive side" (P13). Further, the cue encouraged critical engagement with the AI responses and influenced participants' subsequent queries: "This [cue] helped me a lot because I realized I was only talking about positive things because that is what was output by Perplexity. But I didn't really think about that until I saw this [cue]… So let me prompt it to give me negative things so I can have two perspectives." (P37). Some participants noted that it helped them think more deeply about what to search next: "I looked at what I might be missing from the current understanding, and I was like, okay, I have the key [arguments] over why this is good and why this is bad. And I suppose the only question I have is how do I make more of the good and make less of the bad?" (P23). Only a few participants found it unhelpful. One of them mentioned that it was due to lack of trust in Perplexity's ability to generate any new perspectives, as they believed it only provides repetitive responses. Another suggested the cue could be more useful if alternative perspectives were explicitly provided.

The **Consolidation cue** served its intended purpose by helping participants reflect on what they had learned and synthesize a large amount of information into key points in their own words: "I think this [cue] was really helpful, especially I had so much information that it just made it easier to put just few points down." (P13). Participants noted that this process strengthened their understanding, such as by reviewing their notes: "I think that was helpful just to make sure that I actually understood what I was reading instead of just writing it down and kind of just not passing it to my own memory." (P39), "I thought it was pretty helpful in just evaluating all my notes" (P11). Those who found it less helpful cited personal preferences, or a preference to continue searching rather than summarizing, given the time constraints of the task: "I know there's data that shows [summarizing] is helpful, but it's just not for me" (P3). One participant requested a summary of the topic from Perplexity and subsequently summarized the AI-generated response in their own words.

Overall, participants' perceptions of the cues and their influence on their thought processes were largely positive. Most participants found the cues helpful in prompting deeper engagement, encouraging consideration of multiple perspectives, and helping them search in the right direction. One participant noted, "I think it helps me think in the right direction because otherwise I would just be very aimlessly searching for some things" (P27), while another remarked, "I think without them I would've maybe just stuck to more surface level information… But since it kept giving me [cues], it made me dive deeper and want to keep writing" (P31). However, a few participants reported that the cues were not particularly useful, attributing this to high confidence in their own thinking and search strategies: "I tend to be the person who thinks really deeply and already have a structure of what I wanted, and so I didn't really see anything that would make me deviate from the structure I already had in my head" (P29). A few other participants reported that certain cues disrupted their thought process, leading them to disregard them: "Sometimes they pop up when I'm already writing something, so I'm focused on a thought, and I might disregard the [cue] a little bit because I'm already doing something else" (P33).

*Perceived Helpfulness of Cues (RQ3)*
Regarding RQ3, which considers how the 20 *Cues* condition participants perceived the helpfulness of metacognitive prompts in GenAI search, Figure 3 presents the distribution of participants' ratings on the perceived helpfulness of each type of cue, as reported in the post-task survey. Participants rated the Broadening perspective cue as the most helpful, followed by the Consolidation and Comprehension cues. The Orienting cue was perceived as the least helpful. Notably, the highest proportion of participants did not recall seeing the Comprehension cue; however, among those who did, most found it at least somewhat helpful. This could be explained by the timing of this cue's delivery, which was during pauses in searching or note-taking—when participants could have been engaged in deep thought, potentially causing them to overlook it. The delivery of other types of cues during potential breaks in deep thinking—such as while a GenAI response was getting generated or just before typing a new query—could have been effective in capturing participants' attention.

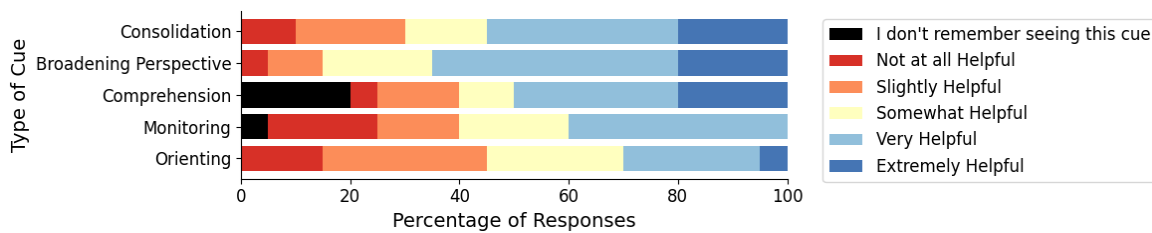

**Figure 3. Perceived helpfulness of each type of cue**

**DISCUSSION**
Through a between-subjects user study with 40 university students, we examined how metacognitive prompting through cues influences search behaviors and critical thinking during GenAI tool use. We found that the cues led to more active engagement, as evidenced by significantly more topics searched ($p=0.006$) and greater persistent inquiry ($p=0.047$), which refers to inquiring about any idea through relevant follow-up queries and is a key component of critical thinking (Dewey, 1910). Further, the cues led to marginally significant increase in search query count ($p=0.053$), including those aimed at comprehension ($p=0.068$) and critical inquiry ($p=0.100$). Retrospective think-aloud interviews confirmed that each cue generally fulfilled its intended purpose. In some cases, the cues also had unexpected but positive effects. For example, the comprehension cue, designed to help users identify gaps in understanding and encourage deeper engagement, prompted one participant to take detailed notes for future reference, demonstrating proactive reflection. The broadening perspective cue, intended to encourage the consideration of alternative or overlooked perspectives, led some participants to critically assess the AI responses upon noticing biased coverage of perspectives by Perplexity.

However, a few participants responded to the cues in unintended ways. For instance, one participant responded to the Consolidation cue, which encouraged them to summarize what they had learned, by asking Perplexity to summarize the topic instead. Others reported that certain cues were less helpful due to confusion about their intended purpose or a belief that they lacked relevance or value. Furthermore, participants with high confidence in their topic knowledge or search skills, and fixed perceptions of Perplexity's performance—e.g., that it generates repetitive responses or lacks memory of prior queries—may have perceived the cues as less impactful.

Prior research has shown that metacognitive prompts are effective for those who possess metacognitive skills but may not be able to apply them spontaneously (Bannert & Mengelkamp, 2013). One such skill is metacognitive flexibility—the ability to adaptively shift cognitive strategies when encountering new information, such as realizing that a current strategy isn't effective (Cañas et al., 2007). Effective use of GenAI tools requires iterating on prompting strategies with respect to the GenAI tool. This, in turn, involves evaluating AI responses, calibrating one's confidence in prompting skills, and separating these from assumptions about the tool's capabilities (Tankelevitch et al. 2023). The observation that some participants attributed unsatisfactory responses to Perplexity's potential limitations, rather than reflecting on the quality of their own queries, suggests limited metacognitive flexibility. Similarly, high confidence in prior knowledge of the topic could have hindered critical evaluation of AI responses and reduced the perceived helpfulness of the Monitoring and Comprehension cues.

To improve the effectiveness of metacognitive prompts, we offer some recommendations and insights. To educators, we recommend: (i) Communicate the purpose and value of each type of cue in the context of GenAI search through pre-search tutorials, as recommended in prior work (Guo, 2021). This is particularly relevant given the lack of significant differences in perceived helpfulness of the note-taking tool across conditions and concerns that some participants found the cues disruptive; (ii) Combine metacognitive prompts with AI literacy interventions that focus on GenAI's capabilities and limitations. Provide guidance on crafting search queries effectively, such as by specifying context and adding details to generate better AI responses (Oppenlaender et al., 2024); (iii) For users with underdeveloped metacognitive skills, future work should explore more direct forms of support, such as metacognitive instructions that provide clear initial structures for learners to know what they need to do and explicit guidance on self-monitoring and regulation during GenAI tool use (Alberto & Troutman, 2013; Zhou & Lam, 2019).

For developers and designers, we note the following: (i) Some participants struggled to act on the cues due to confusion about their intent, which was an expected outcome, as the cues were intentionally high-level to give users autonomy in how to respond. Leveraging GenAI to generate optional, more specific guidance on request, similar to incremental hint generation in educational contexts (Antonucci et al., 2015), can enhance the cues' effectiveness; (ii) As we identified measurable indicators of critical thinking such as persistent inquiry, independent thinking, and source engagement, future work should focus on developing cues tailored to these specific measures. Moreover, leveraging AI to capture these measures and automatically deliver cues based on user behaviors would be a promising direction. For instance, a cue to promote persistent inquiry can be sent when a user has not asked follow-up questions for an extended period. The Comprehension cue, which currently combined assessing comprehension, promoting source engagement, and encouraging persistent questioning, could be separated into three distinct cues; (iii) Future work should explore seamless integration of cues into users' workflows, including optimal placement within the interface and timing of delivery. To better capture users' attention, cues could be sent when they are less likely engaged in deep thinking, such as when a GenAI response is being generated or when they transition from note-taking to typing a new query.

## LIMITATIONS
This study was conducted with students from a single university, though they represented diverse majors, partially addressing concerns about generalizability. Participants were asked to search a single topic to allow in-depth exploration without fatigue and to facilitate comparison across participants. However, the chosen topic may have influenced search and learning behaviors. Future studies should examine the impact of metacognitive prompts for a broader range of topics. The study conditions may not fully reflect authentic learning environments, further limiting generalizability. The Wizard-of-Oz method, while useful for controlled exploration of metacognitive prompts' design and development, does not represent a fully autonomous system. Finally, the relatively small sample size suggests the need for future research with a larger sample of students.

## CONCLUSION
This study examined the effectiveness of metacognitive prompts in enhancing students' critical thinking while searching with GenAI tools, using a multi-method approach that involved user studies, surveys, and retrospective think-aloud interviews. Our findings provide empirical support for the role of metacognitive prompts in fostering active engagement and critical thinking during GenAI-supported search. Specifically, we introduced five types of metacognitive prompts—Orienting, Monitoring, Comprehension, Broadening Perspective, and Consolidation—and identified three quantifiable indicators of critical thinking in this context: persistent inquiry, independent thinking, and source engagement. Although this study focused on a single GenAI tool, the findings broadly offer implications on how metacognitive prompts can be used to support critical thinking in human-AI interactions. Our analysis indicates that the effects of these prompts may vary based on students' metacognitive flexibility. To support a broader range of students, we provided insights to improve the effectiveness of metacognitive prompts, such as explicitly communicating their value and intent and pairing metacognitive prompts with AI literacy instruction. Furthermore, this study contributes to the Search as Learning (SAL) research community by identifying specific behavioral indicators of critical thinking during search interactions with GenAI tools. Future research should explore a broader range of critical thinking measures and investigate the long-term effects of metacognitive prompting on learning outcomes such as knowledge retention and transfer.

**GENERATIVE AI USE**

We used ChatGPT to improve the clarity and coherence of this paper. We evaluated the output by reviewing, verifying, and revising all the generated content. Study participants were instructed to use Perplexity.ai, as detailed above. The authors assume all responsibility for the content of this submission.

**AUTHOR ATTRIBUTION**

First Author: conceptualization, project administration, methodology, data curation, formal analysis, investigation, visualization, supervision, writing – original draft; Second Author: methodology, data curation, formal analysis, investigation, visualization, writing – original draft; Third Author: project administration, funding acquisition, supervision, validation, writing – review and editing.

**ACKNOWLEDGMENTS**

We thank Yu-Ting (Dara) Chang for developing the note-taking interfaces and the Wizard view interface for delivering the cues. We also thank Jacek Gwizdka and all the pilot study participants for their feedback.

**REFERENCES**

Alberto, P., & Troutman, A. C. (2013). *Applied behavior analysis for teachers* (9th ed.). Pearson.

Amer, E., & Elboghdadly, T. (2024). The end of the search engine era and the rise of generative AI: A paradigm shift in information retrieval. In *2024 International Mobile, Intelligent, and Ubiquitous Computing Conference (MIUCC)* (pp. 374–379).

Antonucci, P., Estler, C., Nikolić, D., Piccioni, M., & Meyer, B. (2015). An incremental hint system for automated programming assignments. In *Proceedings of the 2015 ACM Conference on Innovation and Technology in Computer Science Education*.

Bannert, M., & Mengelkamp, C. (2013). Scaffolding hypermedia learning through metacognitive prompts. In R. Azevedo & V. Aleven (Eds.), *International handbook of metacognition and learning technologies* (pp. 171–186). Springer.

Berardi-Coletta, B., Buyer, L. S., Dominowski, R. L., & Rellinger, E. R. (1995). Metacognition and problem solving: A process-oriented approach. *Journal of Experimental Psychology: Learning, Memory, and Cognition*, *21*(1), 205–223.

Beyer, B. K. (1984). Improving thinking skills--defining the problem. *Phi Delta Kappan*, *65*(7), 486–490.

Cañas, J. J., Antolí, A., Fajardo, I., & Salmerón, L. (2005). Cognitive inflexibility and the development and use of strategies for solving complex dynamic problems: Effects of different types of training. *Theoretical Issues in Ergonomics Science*, *6*(1), 95–108.

Cinelli, M., De Francisci Morales, G., Galeazzi, A., Quattrociocchi, W., & Starnini, M. (2021). The echo chamber effect on social media. *Proceedings of the National Academy of Sciences*, *118*(9), e2023301118.

Collins-Thompson, K., Rieh, S. Y., Haynes, C. C., & Syed, R. (2016). Assessing learning outcomes in web search: A comparison of tasks and query strategies. In *Proceedings of the 2016 ACM on Conference on Human Information Interaction and Retrieval* (pp. 163–172). ACM.

Dewey, J. (1910). *How we think*. D.C. Heath & Co.

Efklides, A. (2008). Metacognition: Defining its facets and levels of functioning in relation to self-regulation and co-regulation. *European Psychologist*, *13*(4), 277–287.

Eickhoff, C., Gwizdka, J., Hauff, C., & He, J. (2017). Introduction to the special issue on search as learning. *Information Retrieval Journal*, *20*, 399–402.

Elling, S., Lentz, L., & de Jong, M. (2011). Retrospective think-aloud method: Using eye movements as an extra cue for participants' verbalizations. In *Proceedings of the SIGCHI Conference on Human Factors in Computing Systems* (pp. 1161–1170).

Fan, Y., Tang, L., Le, H., Shen, K., Tan, S., Zhao, Y., Shen, Y., Li, X., & Gašević, D. (2025). Beware of metacognitive laziness: Effects of generative artificial intelligence on learning motivation, processes, and performance. *British Journal of Educational Technology*, *56*(2), 489–530.

Flavell, J. H. (1979). Metacognition and cognitive monitoring: A new area of cognitive–developmental inquiry. *American Psychologist*, *34*(10), 906–911.

Gerlich, M. (2025). AI tools in society: Impacts on cognitive offloading and the future of critical thinking. *Societies*, *15*(1), 6.

Guo, L. (2022). Using metacognitive prompts to enhance self-regulated learning and learning outcomes: A meta-analysis of experimental studies in computer-based learning environments. *Journal of Computer Assisted Learning*, *38*(3), 811–832.

Halpern, D. F. (2013). *Thought and knowledge: An introduction to critical thinking* (5th ed.). Psychology Press.

Hansen, P., & Rieh, S. Y. (2016). Recent advances on searching as learning: An introduction to the special issue. *Journal of Information Science*, *42*(1), 3–6.

Hwang, G.-J., & Kuo, F.-R. (2011). An information-summarising instruction strategy for improving the web-based problem solving abilities of students. *Australasian Journal of Educational Technology*, *27*(2).

Jansen, B. J., Booth, D., & Smith, B. (2009). Using the taxonomy of cognitive learning to model online searching. *Information Processing & Management*, *45*(6), 643–663.

Jansen, B. J., Smith, B., & Booth, D. L. (2007). Understanding web search via a learning paradigm. In *Proceedings of the 16th International Conference on World Wide Web* (pp. 1207–1208). ACM.

Kidd, C., & Birhane, A. (2023). How AI can distort human beliefs. *Science*, *380*(6651), 1222–1223.

Ku, K. Y. L., & Ho, I. T. (2010). Metacognitive strategies that enhance critical thinking. *Metacognition and Learning*, *5*(3), 251–267.

Leeder, C., & Shah, C. (2016). Practicing critical evaluation of online sources improves student search behavior. *The Journal of Academic Librarianship*, *42*, 459–468.

Lin, X. (2001). Designing metacognitive activities. *Educational Technology Research and Development*, *49*(2), 23–40.

Lin, X., Hmelo, C., Kinzer, C. K., & Secules, T. J. (1999). Designing technology to support reflection. *Educational Technology Research and Development*, *47*(3), 43–62.

Lin, X., & Lehman, J. D. (1999). Supporting learning of variable control in a computer-based biology environment: Effects of prompting college students to reflect on their own thinking. *Journal of Research in Science Teaching*, *36*(7), 837–858.

Magno, C. (2010). The role of metacognitive skills in developing critical thinking. *Metacognition and Learning*, *5*(2), 137–156.

Makany, T., Kemp, J., & Dror, I. E. (2009). Optimising the use of note-taking as an external cognitive aid for increasing learning. *British Journal of Educational Technology*, *40*(4), 619–635.

Maulsby, D., Greenberg, S., & Mander, R. (1993). Prototyping an intelligent agent through wizard of oz. In *Proceedings of the INTERACT '93 and CHI '93 Conference on Human Factors in Computing Systems* (pp. 277–284). Association for Computing Machinery.

Mayer, R. E., & Goodchild, F. M. (1990). *The critical thinker: Thinking and learning strategies for psychology students*. Wm. C. Brown.

Nelson, T. O., & Narens, L. (1990). Metamemory: A theoretical framework and new findings. In *Psychology of learning and motivation* (Vol. 26, pp. 125–173). Elsevier.

Niedbał, R., Sokołowski, A., & Wrzalik, A. (2023). Students' use of the artificial intelligence language model in their learning process. *Procedia Computer Science*, *225*, 3059–3066.

Odgers, C. L., & Jensen, M. R. (2020). Annual research review: Adolescent mental health in the digital age: Facts, fears, and future directions. *Journal of Child Psychology and Psychiatry*, *61*(3), 336–348.

Oppenlaender, J., Linder, R., & Silvennoinen, J. (2024). Prompting AI art: An investigation into the creative skill of prompt engineering. *International Journal of Human–Computer Interaction*, 1–23.

Pritschet, L., Powell, D., & Horne, Z. (2016). Marginally significant effects as evidence for hypotheses: Changing attitudes over four decades. *Psychological Science*, *27*(7), 1036–1042.

Rieh, S. Y., Gwizdka, J., Freund, L., & Collins-Thompson, K. (2014). Searching as learning: Novel measures for information interaction research. *Proceedings of the American Society for Information Science and Technology*, *51*(1), 1–4.

Rieh, S. Y., Collins-Thompson, K., Hansen, P., & Lee, H.-J. (2016). Towards searching as a learning process: A review of current perspectives and future directions. *Journal of Information Science*, *42*(1), 19–34.

Risko, E. F., & Gilbert, S. J. (2016). Cognitive offloading. *Trends in Cognitive Sciences*, *20*(9), 676–688.

Roy, N., Valle Torre, M., Gadiraju, U., Maxwell, D., & Hauff, C. (2021). Note the highlight: Incorporating active reading tools in a search as learning environment. In *Proceedings of the 2021 Conference on Human Information Interaction and Retrieval* (pp. 229–238).

Schraw, G., & Dennison, R. S. (1994). Assessing metacognitive awareness. *Contemporary Educational Psychology*, *19*(4), 460–475.

Shah, C., & Bender, E. M. (2024). Envisioning information access systems: What makes for good tools and a healthy web? *ACM Transactions on the Web*, *18*(3), 33:1–33:24.

Sharma, N., Liao, Q. V., & Xiao, Z. (2024). Generative echo chamber? Effect of LLM-powered search systems on diverse information seeking. In *Proceedings of the 2024 CHI Conference on Human Factors in Computing Systems* (pp. 1–17).

Singh, A., Taneja, K., Guan, Z., & Ghosh, A. (2025). Protecting human cognition in the age of AI. In *Tools for Thought Workshop at the 2025 CHI Conference on Human Factors in Computing Systems*.

Solaiman, I., Talat, Z., Agnew, W., Ahmad, L., Baker, D., Blodgett, S. L., Chen, C., Clark, K., Curry, A., Ecoffet, A., Elkins, K., Evans, O., Ganguli, D., Gerigk, L., Hernandez, D., Hilton, J., Hume, T., Jackson, S., Jain, N., ... Wu, S. (2023). Evaluating the social impact of generative AI systems in systems and society. *arXiv preprint arXiv:2306.05949*.

Stadtler, M., & Bromme, R. (2007). Dealing with multiple documents on the WWW: The role of metacognition in the formation of documents models. *International Journal of Computer-Supported Collaborative Learning*, *2*, 191–210.

Stadtler, M., & Bromme, R. (2008). Effects of the metacognitive computer-tool Met.a.Ware on the web search of laypersons. *Computers in Human Behavior*, *24*(3), 716–737.

Tankelevitch, L., Kewenig, V., Simkute, A., Scott, A. E., Sarkar, A., Sellen, A., & Rintel, S. (2024). The metacognitive demands and opportunities of generative AI. In *Proceedings of the CHI Conference on Human Factors in Computing Systems* (pp. 1–24).

*Teen and young adult perspectives on generative AI*. (2024). Common Sense Media. https://www.commonsensemedia.org/sites/default/files/research/report/teen-and-young-adult-perspectives-on-generative-ai.pdf

Urgo, K., & Arguello, J. (2025). Search as learning. *Foundations and Trends® in Information Retrieval*, *19*(4), 365–556.

Vakkari, P. (2016). Searching as learning: A systematization based on literature. *Journal of Information Science*, *42*(1), 7–18.

Venkit, P. N., Laban, P., Zhou, Y., Mao, Y., & Wu, C.-S. (2024). Search engines in an AI era: The false promise of factual and verifiable source-cited responses. *arXiv preprint arXiv:2410.22349*.

Weir, J. P. (2005). Quantifying test-retest reliability using the intraclass correlation coefficient and the SEM. *The Journal of Strength & Conditioning Research*, *19*(1), 231–240.

Winne, P. H. (2018). Cognition and metacognition within self-regulated learning. In D. H. Schunk & J. A. Greene (Eds.), *Handbook of self-regulation of learning and performance* (2nd ed., pp. 36–48). Routledge/Taylor & Francis Group.

Zamani, H., Trippas, J. R., Dalton, J., & Radlinski, F. (2023). Conversational information seeking. *Foundations and Trends® in Information Retrieval*, *17*(3-4), 244–456.

Zhou, M., & Lam, K. K. L. (2019). Metacognitive scaffolding for online information search in K-12 and higher education settings: A systematic review. *Educational Technology Research and Development*, *67*(6), 1353–1384.

Zhou, T., & Li, S. (2024). Understanding user switch of information seeking: From search engines to generative AI. *Journal of Librarianship and Information Science*, 1–13.

Zimmerman, B. J. (2002). Becoming a self-regulated learner: An overview. *Theory into Practice*, *41*(2), 64–70.

**APPENDIX**

Expanded definitions of measures capturing search behavior and critical thinking.

| Metric (data source) | Definition | Type |
|---|---|---|
| **# Topics** (video recording of search session) | Number of unique and relevant topics and sub-topic, i.e., 'information focus' of one or more queries (Jansen et al., 2007) *excluding* given topic. The following are NOT considered as topics: | Integer |

| | | | |
|---|---|---|---|
| | | - context related to sources<br>- asking for search related guidance<br>- sub-topics of a topic that has already been counted<br><br>Multiple aspects, e.g., 'pros and cons' should be considered individually. In this case, 'pros' and 'cons' should be considered as 2 topics<br>Typically, a search query has one topic unless it consists of multiple sub-queries or multiple aspects as described above. While multiple perspectives count as multiple topics when , however the impact of X on Y, even though it consists of 2 entities, counts as a single topic.<br><br>Examples of search queries, with topics (in bold, different topics in different colors):<br>1. Are there **challenges with AI replacing human qualities** in education?<br>2. **Ethical Considerations** and **Bias**<br>3. How do **AI tutoring systems handle emotional support** for students<br>4. How is **AI being used to "support educators"** specifically? What **AI tools** exist and are currently being used to perform the tasks that are mentioned? | |
| **Cognitive level of queries**<br><br>(video recording of search session) | Number of **unique** and **relevant** queries that fall into either of the two categories: *receptive* vs *critical*. Adding context related to sources or search related guidance to an existing query does NOT count as a unique query. The given topic also does NOT count as a unique query.<br><br>We consider the number of each of the following types of queries:<br><br>- **Receptive:** *remembering or understanding information*<br>  Queries typically aimed towards understanding or comprehension.<br><br>  Examples:<br>  1. Give me analysis of how AI is impacting education based on widely known knowledge and provide me with pros and cons.<br>  2. What are the main challenges educators face when integrating AI into the curriculum.<br>  3. What's currently being done with ai in education<br><br>- **Critical:** *applying, analyzing, or evaluating information*<br>  Queries about **opinions**, **arguments**, gathering **evidence**, **predicting**, or **digging deeper** into a topic. A query about *purposefully* contrasting multiple perspectives would be critical, however a query about simply understanding multiple entities or perspectives would be receptive. Even if a query is partially critical, it should be counted as a critical query.<br><br>  Examples:<br>  1. Are there more papers saying that the quality of education improved due to AI or more arguing that is has declined?<br>  2. Is there any evidence students are learning critical thinking by assessing the reliability of AI-generated content<br>  3. Should students be allowed to use AI for assignments? | Integer (for each category: receptive and critical) |

| Independent thinking (video recording of search session) | Adding new inputs in notes in addition to information provided by Perplexity and sources, instead of copying/rewriting information without any input of their own. E.g., connecting information from Perplexity to prior knowledge, posing new questions or reflecting on information generated by Perplexity. | 0 (no) 1 (yes) |
|---|---|---|
| Persistent inquiry (video recording of search session) | Inquiry of an idea through querying followed by **relevant follow-up queries**. | 0 (no) 1 (yes) |
| Source Engagement (video recordings of search session and retrospective think aloud interviews) | Gathering information from sources to support claims made by Perplexity, test hypotheses, get more details or improve understanding. Attempting to do so unsuccessfully does not count. | 0 (no) 1 (yes) |